\documentclass{iopart}

\usepackage{iopams}
\usepackage{caption}
\usepackage{subcaption}
\usepackage{graphicx}
\usepackage{xcolor}
\usepackage{makecell}
\usepackage[normalem]{ulem}
\definecolor{Mygrey}{rgb}{0.3 0.3 0.3}

\usepackage{mathrsfs}

\usepackage{hyperref}
\hypersetup{
    colorlinks=true,
    linkcolor=Mygrey,
    urlcolor=Mygrey,
    citecolor=Mygrey
}

\expandafter\let\csname equation*\endcsname\relax
\expandafter\let\csname endequation*\endcsname\relax
\usepackage{amsmath}

\newcommand{\drm}{{\rm d}}

\newcommand{\beq}{\begin{equation}}
\newcommand{\eeq}{\end{equation}}
\newcommand{\bdm}{\begin{displaymath}}
\newcommand{\edm}{\end{displaymath}}

\begin{document}

\title[]
{Machine learning for gravitational-wave detection: surrogate Wiener filtering for the prediction and optimized cancellation of Newtonian noise at Virgo}

\author{F Badaracco$^{1,2}$ and J Harms$^{1,2}$ and A Bertolini$^{3}$ and T Bulik$^{4}$ and I Fiori$^{5}$ and B Idzkowski$^{4}$ and A Kutynia$^{4}$ and K Nikliborc$^{4}$ and F Paoletti$^{6}$ and A Paoli$^{5}$ and L Rei$^{7}$ and M Suchinski$^{4}$}

\vskip 1mm
\address{$^{1}$Gran Sasso Science Institute (GSSI), I-67100 L'Aquila, Italy}
\address{$^{2}$INFN, Laboratori Nazionali del Gran Sasso, I-67100 Assergi, Italy}
\address{$^{3}$Nikhef, Science Park 105, 1098 XG Amsterdam, Netherlands}
\address{$^{4}$Astronomical Observatory, University of Warsaw, Aleje Ujazdowskie 4, 00-478 Warsaw, PL}
\address{$^{5}$European Gravitational Observatory (EGO), I-56021 Cascina, Pisa, Italy}
\address{$^{6}$INFN, Sezione di Pisa, I-56127 Pisa, Italy}
\address{$^{7}$INFN, Sezione di Genova, I-16146 Genova, Italy}

\begin{abstract}
The cancellation of noise from terrestrial gravity fluctuations, also known as Newtonian noise (NN), in  gravitational-wave detectors is a formidable challenge. Gravity fluctuations result from density perturbations associated with environmental fields, e.g., seismic and acoustic fields, which are characterized by complex spatial correlations. Measurements of these fields necessarily provide incomplete information, and the question is how to make optimal use of available information for the design of a noise-cancellation system. In this paper, we present a machine-learning approach to calculate a surrogate model of a Wiener filter. The model is used to calculate optimal configurations of seismometer arrays for a varying number of sensors, which is the missing keystone for the design of NN cancellation systems. The optimization results indicate that efficient noise cancellation can be achieved even for complex seismic fields with relatively few seismometers provided that they are deployed in optimal configurations. In the form presented here, the optimization method can be applied to all current and future gravitational-wave detectors located at the surface and with minor modifications also to future underground detectors.
\end{abstract}


\section{Introduction}
Gravitational-wave (GW) detectors are vulnerable to a variety of environmental disturbances, which can reduce their duty cycle and sensitivity. These disturbances have their origin in seismic, atmospheric, and magnetic fields. Isolation of the suspended test masses and other relevant parts of the system from the environment is essential to the detection of gravitational waves, which cause minuscule changes of distance between test masses. The vacuum system and seismic isolation are the key ingredients to decouple the sensitive parts of the instrument from the environment. The seismic isolation mostly consists of mechanical units such as pendula and cantilevers capable of suppressing the propagation of environmental disturbances to the sensitive parts like the test masses. A so-called active isolation relying on sensors and actuators further improves the isolation performance by suppressing vibrations in the system. 

Fluctuations of the terrestrial gravity field directly couple with the test masses and therefore circumvent the isolation systems leading to Newtonian noise \cite{Har2019}. They are produced by seismic fields \cite{Sau1984,HuTh1998,BeEA1998}, atmospheric fields \cite{Cre2008, FiEA2018}, and also moving and vibrating objects \cite{Har2019}. For the Virgo detector, the dominant contributions to NN are predicted to come from seismic and acoustic fields \cite{Singha2020,FiEA2018}. Low-frequency noise including Newtonian noise has significant impact on parameter estimation of compact-binary GW signals \cite{LyEA2015,ViEv2017,HaEv2019}, and it can also significantly influence the signal-to-noise ratio especially of intermediate-mass black-holes visible mostly through their harmonics excited during the merger \cite{BrSe2007}. Mitigation of NN therefore has an important impact on the science that can be done with GW observations.

New methods to lower environmental noise are being developed and implemented. One approach is to lower disturbances in the environment, which is possible whenever the sources are under human control like pumps and ventilation systems. Another method is the so-called offline subtraction of noise, where data from environmental sensors are passed through filters and subtracted from the detector data. This method has been implemented successfully in LIGO and Virgo, for example, to reduce noise from vibrations of optical tables causing laser-beam jitter. 

A combination of these two methods is also being considered to mitigate NN at Virgo as part of the Advanced Virgo+ detector upgrade. For the NN offline subtraction \cite{Cel2000, CoEA2018a}, the plan is to deploy around 140 seismometers inside the three main experimental halls of the Virgo detector to monitor the seismic field and to estimate the associated gravity fluctuations. Extensive seismic studies are being carried out to characterize the field in terms of its spatial and temporal properties \cite{Tringali2019}. The goal is to understand from the observed properties of the seismic field, especially its two-point correlations, how to deploy the seismometers. 

Optimization of array configurations for NN cancellation has been a difficult challenge, and so far, it was only possible to calculate optimal arrays for simple fields where seismic correlations and the gravity perturbation have a known analytic expression \cite{DHA2012,CoEA2016a}. The main challenges are
\begin{itemize}
\item The optimization involves a large number of variables (2 coordinates per seismometer).
\item Real seismic fields, especially those at the Virgo site where local seismic sources dominate and seismic waves interact with a complex infrastructure \cite{Tringali2019,Singha2020}, cannot be represented by analytic models. 
\item Information about the seismic field obtained by site-characterization measurements with arrays is incomplete.
\item How to systematically and optimally use information about the seismic field for the design of a NN cancellation system has been an open problem so far.
\end{itemize}

In this paper, we present an efficient approach to the optimal design of a NN cancellation system based on observed two-point spatial correlations between seismometers deployed in an array. The solution takes the form of a surrogate Wiener filter with seismometers as input channels, and whose output constitutes an estimate of the gravity fluctuation produced by the seismic field. It can be calculated for an arbitrary number of seismometers with arbitrary positions on the surface. The method incorporates kriging --- as Gaussian process regression is sometimes called when spatial correlations are involved --- and simple interpolation, and is devised to address computational limitations (the optimization procedure still requires a computer cluster to obtain robust optimization results). 

A summary of the seismic experiment at Virgo whose data were used to calculate the surrogate model is given in section \ref{sec:Seismometers}. In section \ref{sec:surrWF}, we describe the construction of the Wiener-filter surrogate model and some of its properties. The optimization results are presented in section \ref{sec:opt}.

\section{Instruments and positioning}
\label{sec:Seismometers}
\begin{figure}[ht!]
\centering
\includegraphics[width = 0.8\textwidth]{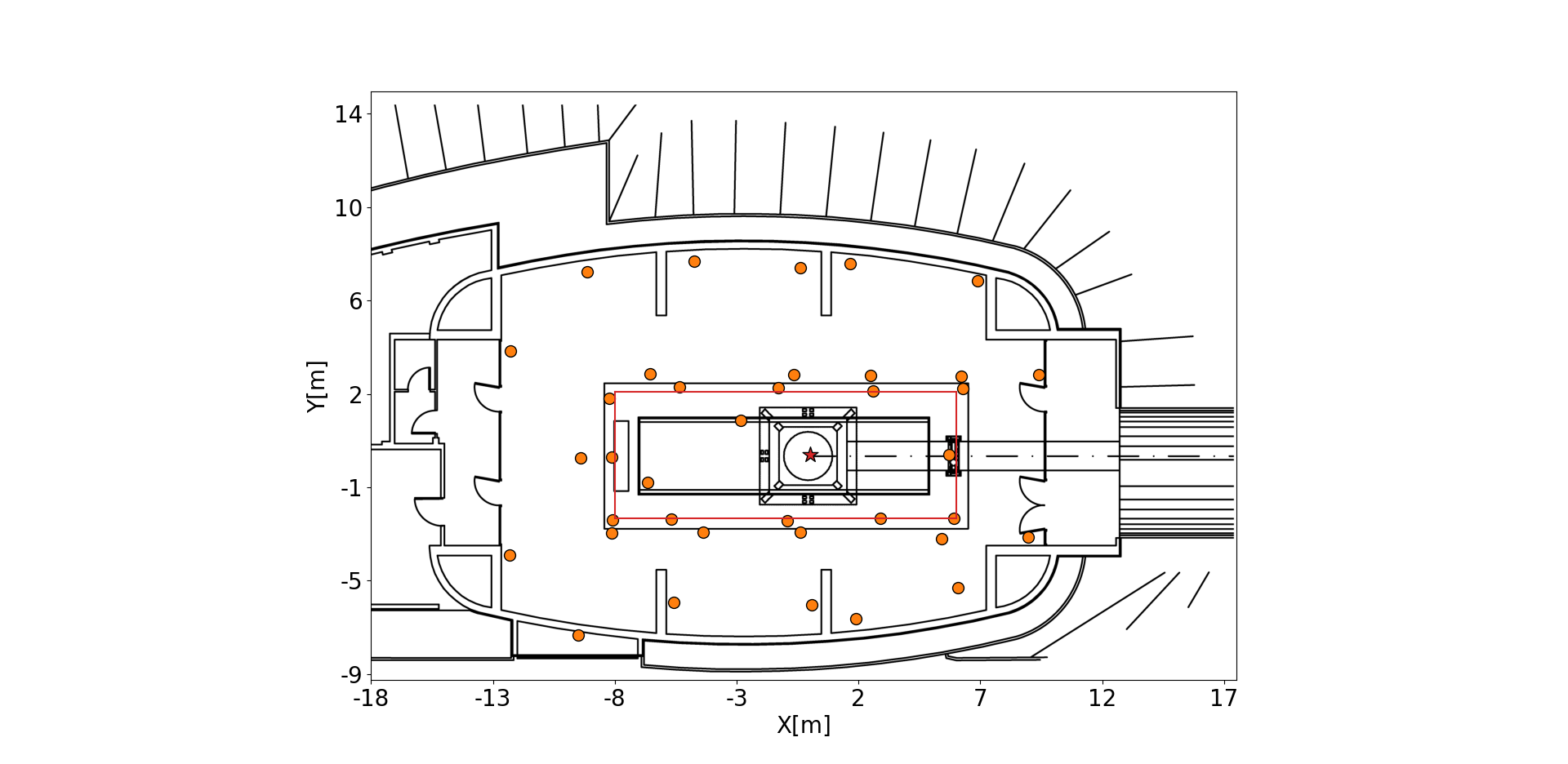}
\caption{Plan of the WEB seismic array.}
\label{fig:array}
\end{figure}

In this work we used the data collected from $38$ seismometers placed in the Virgo West-End Building (WEB) \cite{Tringali2019}. The seismometer positions are shown in figure \ref{fig:array}.
The WEB hosts one of Virgo's suspended test masses (red star in figure \ref{fig:array}). It has a complicated structure: the ground is not homogeneous and consists of a basement under the test mass whose floor is $3.5\,$m below surface (horizontal extent marked by a red rectangle). The ceiling and walls of the basement are disconnected by a thin gap from the main building floor. The entire structure supported by the basement is called tower platform. It is anchored with $52 \,$m deep pillars to the hard rock beneath the soft soil to prevent it from sinking since it carries the vacuum chamber of the seismic isolation and test-mass suspension system. 

We place the origin of our coordinate system at the surface level right below the test mass. The basement extents from -8\,m to 6\,m along $x$, and $\pm 2.6\,$m along $y$ (see Singha et al for more details \cite{Singha2020}). Of the $38$ seismometers placed in the WEB, $15$ were placed on the tower platform (of which only two were on the floor of the basement) and $23$ on the building floor. One seismometer was discarded since it was deployed on a metal sheet that forms part of the ceiling of the basement, and whose vibrations are uncorrelated with the seismic field. Also, the metal sheet is too thin to contribute significantly to NN. 

The seismometers are 5\,Hz geophones with low-noise readout nodes. For a detailed description of the instrument characteristics and their deployment, see \cite{Tringali2019}. The data used here were taken during one hour between the 00:00 and the 01:00 of February 5, 2018, i.e., a time without human activity in the building and therefore representative of the situation during Virgo observation runs. 

\section{Surrogate model of the Wiener filter}
\label{sec:surrWF}
For noise cancellation, data from witness sensors is passed through a filter and its output subtracted  from a target channel. Normally, the goal is to reduce the variance of the target time series. For the cancellation of stationary noise, the Wiener filter is known to minimize the variance of the residual data \cite{BHC2008}, and they were therefore proposed for NN cancellation in GW detectors \cite{Cel2000}. 

When the source of disturbance is an entire field, as for NN, then effective monitoring of the field becomes the main challenge in the design of a noise-cancellation system. Effective monitoring can be achieved by choosing effective types of seismic sensors \cite{HaVe2016}, and by optimal sensor positioning \cite{DHA2012,CoEA2016a,BaHa2019}. Until now, determining the optimal sensor locations based on correlation measurements of the seismic field has been an open problem.

As described in section \ref{sec:Seismometers}, observations of vertical seismic surface displacement $\xi$ are used here. In frequency domain, a Wiener filter $\vec{W}(\omega)$ is multiplied to the Fourier amplitudes $\vec{\xi}(\omega)$ of $N$ seismic sensors to obtain a NN estimate (here in units of test-mass acceleration):
\beq
\delta\hat a(\omega) = \vec{W}(\omega)\cdot\vec{\xi}(\omega)
\eeq
The Wiener filter takes the form
\beq
\vec W = \langle {\vec\xi}^{\,\dagger} \delta a\rangle\cdot\langle {\vec\xi}\,\vec\xi^{\,\dagger}\rangle^{-1},
\eeq
where $\vec{C}_{\rm SN}\equiv\langle {\vec\xi}\, \delta a^*\rangle$ are the cross spectral densities (CPSDs) between the $N$ seismometers and the test-mass acceleration, and $\mathbf{C}_{\rm SS}\equiv\langle {\vec\xi}\,\vec\xi^{\,\dagger}\rangle$ the cross-spectral density matrix between all seismometers. The correlations $\vec{C}_{\rm SN}$ depend on locations of the test mass and seismometers, and with the test-mass location considered fixed, is a function of two parameters per seismometer (its two horizontal coordinates). Correspondingly, the correlations forming the matrix $\mathbf{C}_{\rm SS}$ depend on the coordinates of a pair seismometers, i.e., four variables.

The Wiener filter minimizes the residual noise for a given configuration of the seismic array. Relative to the NN spectral density $C_{\rm NN}\equiv\langle |\delta a|^2\rangle$, the relative residual can be written as \cite{Cel2000}
\begin{equation}
R(\omega) 	= 1 - \frac{\vec{C}^\dagger_{\rm SN}\left(\omega\right)\cdot\left(\mathbf{C}_{\rm SS}(\omega)\right)^{-1}\cdot \vec{C}_{\rm SN}(\omega)}{C_{\rm NN}(\omega)}
\label{eq:res1}
\end{equation}
The aim is now to minimize $R$ by finding the optimal locations of the $N$ seismometers. This can be achieved by maximizing the numerator in the last equation, since $C_{\rm NN}$ is a constant. In past studies, analytic models were used for $C_{\rm SN}$ and $\mathbf{C}_{\rm SS}$ representing simplified, i.e., isotropic, homogeneous seismic fields. The field observed at the Virgo site is very complicated and cannot be represented by any analytic model \cite{Tringali2019}. The question arises how to make best use of the information we have about the seismic field to estimate the optimal array configuration.

\begin{figure}[ht!]
\centering
\includegraphics[width=0.9\textwidth]{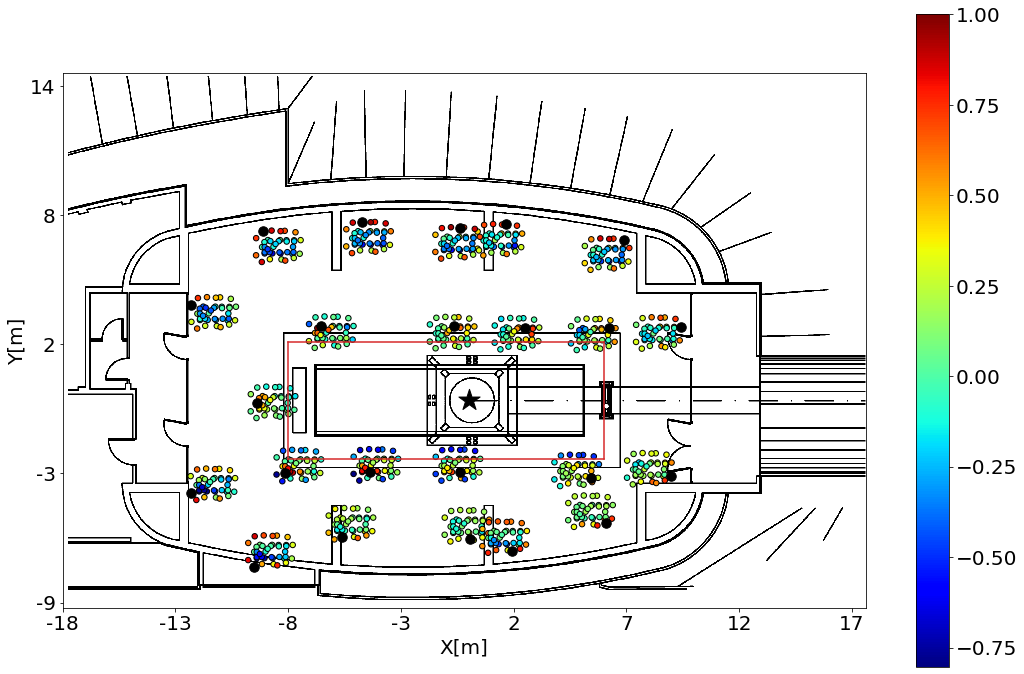}
\includegraphics[width=0.9\textwidth]{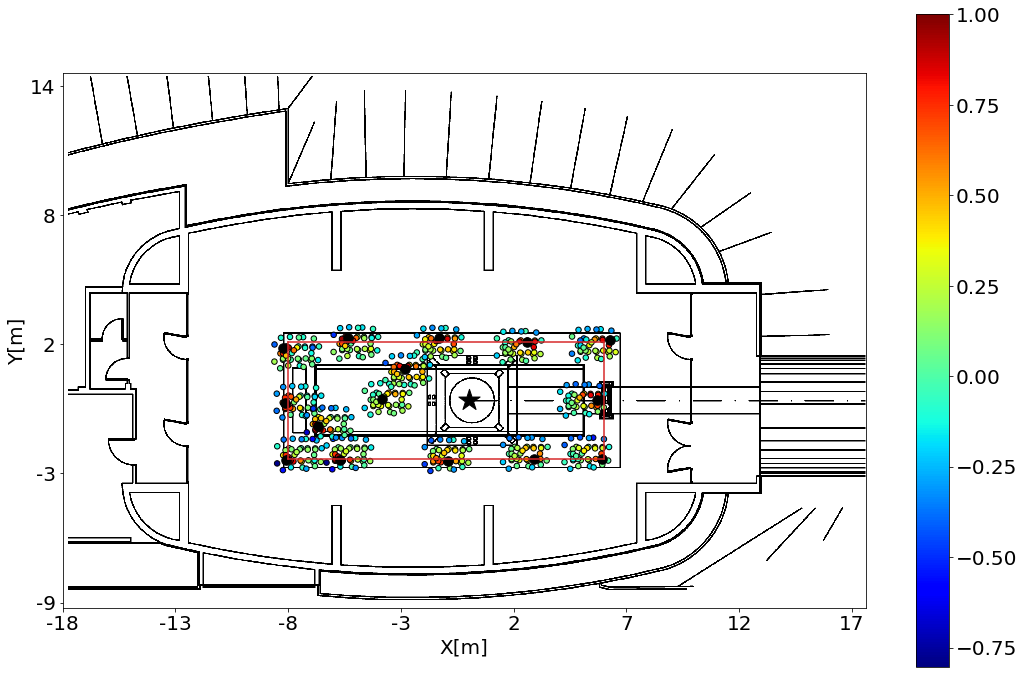}
\caption{Normalized cross-spectral densities (coherence) between all possible pairs of seismometers at 15\,Hz. The correlation values are shown in a down-scaled array configuration where the reference seismometer is marked as black dot. (top) reference sensors on main building floor. (bottom) reference sensors on tower platform.}
\label{fig:corr}
\end{figure}
\begin{figure}[ht!]
\centering
\includegraphics[width=0.8\textwidth]{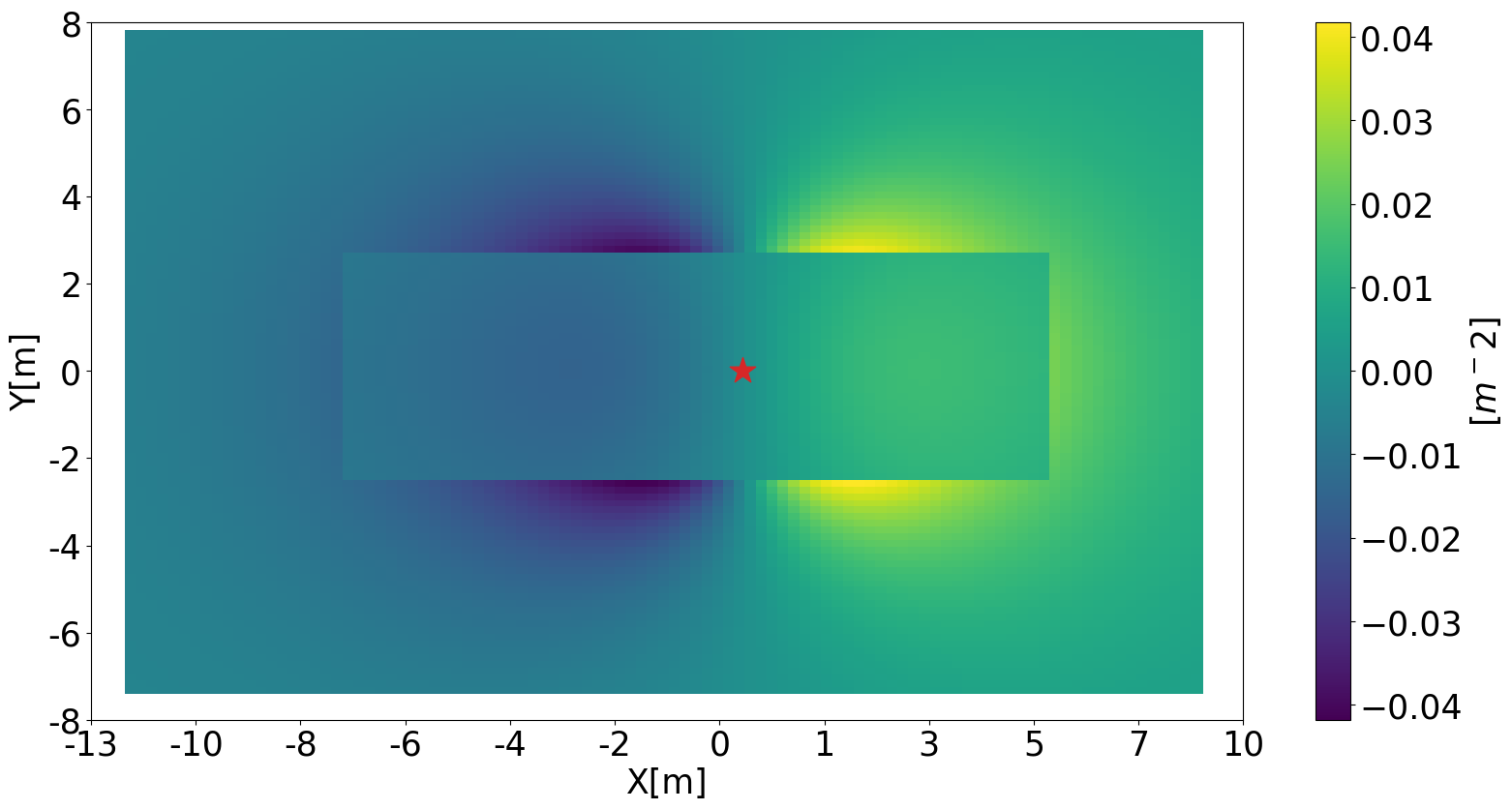}
\caption{Kernel of gravitational coupling between seismic field and NN from Rayleigh waves or surface displacement.}
\label{fig:kernel}
\end{figure}

The approach taken here is to construct a surrogate model of the Wiener filter for an arbitrary number of seismometers, and to use it for the calculation of optimal arrays. Since Virgo's sensitivity is not yet good enough to observe NN, one needs to provide a model of the correlations $C_{\rm SN}$. Assuming that the dominant seismic displacement is produced by Rayleigh waves, or more generally, that NN contributions from surface displacement are dominating over contributions from (de)compression of the ground medium, the required model in terms of seismic correlations reads \cite{CoEA2016a}:
\begin{equation}
	\begin{aligned}
	C_{\rm SN}(\vec r, \vec r_0) &= \mathscr{C}\int C_{\rm SS}(\vec r, \vec r_1)\frac{x_0 - x_1}{(h(x_1,y_1)^2 + |\vec r_1-\vec r_0|^2)^{3/2}}\drm x_1 \drm y_1 \\
	& =\mathscr{C}\int C_{\rm SS}(\vec r, \vec r_1)\mathcal{K}(\vec r_1,\vec r_0)\drm x_1 \drm y_1
	\end{aligned}
\label{eq:Csn}
\end{equation}
where $\vec r_0=(x_0,y_0)$ is the position of the test mass, and $\mathscr{C}$ is a constant not relevant here since it cancels out in the residual $R$ in equation (\ref{eq:res1}). The height $h$ is the vertical distance between the test mass and a surface point ($x_1, y_1$). Due to the tower basement, it has two values, 1.5\,m and 5\,m, for points on the main building floor and on the basement floor, respectively. The kernel $\mathcal K$ links the seismic correlations with NN from Rayleigh waves or surface displacement. Its values are shown in figure \ref{fig:kernel}.

We point out that contributions from the normal displacement of basement walls, despite the fact that they can be included in equation (\ref{eq:Csn}), cannot be considered in our analysis since no sensors were installed on the basement walls.  

For the surrogate model of the Wiener filter, we need $C_{\rm SS}$ as a continuous function of the coordinates of two surface points. This can in principle be done by simple interpolation, but one can achieve much better results with a Bayesian approach, i.e., Gaussian-process regression (GPR), which allows us to implement priors for the correlations, to quantify the errors of inferred correlation values, and also makes it possible to extend the analysis to a region beyond the convex envelope of an array \cite{jones2001taxonomy,forrester2008engineering}. The last point is convenient for technical reasons since we can use a rectangular area for our analysis. 

As shown in figure \ref{fig:corr}, with $N=37$ sensors, we have $37^2$ samples of $C_{\rm SS}$, which is not enough to infer with sufficient accuracy the values of $C_{\rm SS}$ on its 4D domain even with a GPR. In numbers, considering the minimum rectangle containing all the $37$ seismometers, $21.72 \,$m along x and $15.36 \,$m along y, we obtain a 4D volume of $10^5 \,$m$^4$. We can calculate a dimensionless sample density by dividing the volume of $37^2$ hyperspheres of radius $1\,$m by $10^5 \,$m$^4$, which yields the small value $n_{\rm 4D} = 0.06$. The solution to this problem is to apply the GPR to the field of Fourier amplitudes of $\xi$, which is a function of only two coordinates. This increases the sample density to $n_{\rm 2D} = 0.35$. The Fourier amplitudes are inferred on a regular 2D grid of $30^2$ surface points. Each Fourier amplitude has been calculated by means of discrete Fourier transforms using time segments of duration $T$, so that the two-point spatial correlation is given by the average over the Fourier fields of all time segments,
\begin{equation}\label{eq:cpsd}
C_{\rm SS}(\vec r_i, \vec r_j, \omega) = \frac{1}{T}\langle \tilde{\xi}_i(\omega)\tilde{\xi}^*_j(\omega)\rangle,
\end{equation}
where $\tilde{\xi}_i,\,\tilde{\xi}_j$ are the Fourier amplitudes at points $\vec r_i, \vec r_j$. Therefore, with the GPR, we achieve an evaluation of the seismic correlation at $30^4$ points of its 4D domain. The sample density is still not high enough to directly calculate optimal arrays. Another step needs to be added to achieve computationally efficient dense sampling of the seismic correlation as required for array optimization on a continuous coordinate space.

\section{Array optimization}
\label{sec:opt}
Using GPR to estimate $C_{\rm SS}$ is, in principle, enough to run the optimization process. In practice, it is too costly from the computational point of view: it takes too much time evaluating the integral in equation (\ref{eq:cpsd}) through GPR for every iteration of the optimization algorithm. What we have done instead is to use a coarse-grained representation of seismic correlations obtained by GPR (the $30^4$ grid mentioned in section \ref{sec:surrWF}) and then to perform a regular grid linear interpolation to evaluate $C_{\rm SS}$ on arbitrary points as requested by the integration in equation (\ref{eq:Csn}) and optimization algorithm. For the evaluation of the 2D integral in $\vec{C}_{\rm SN}$, we used Simpson's method \cite{press2007numerical}. 

\begin{figure}[ht!]
	\centering
		\includegraphics[width=0.49\textwidth]{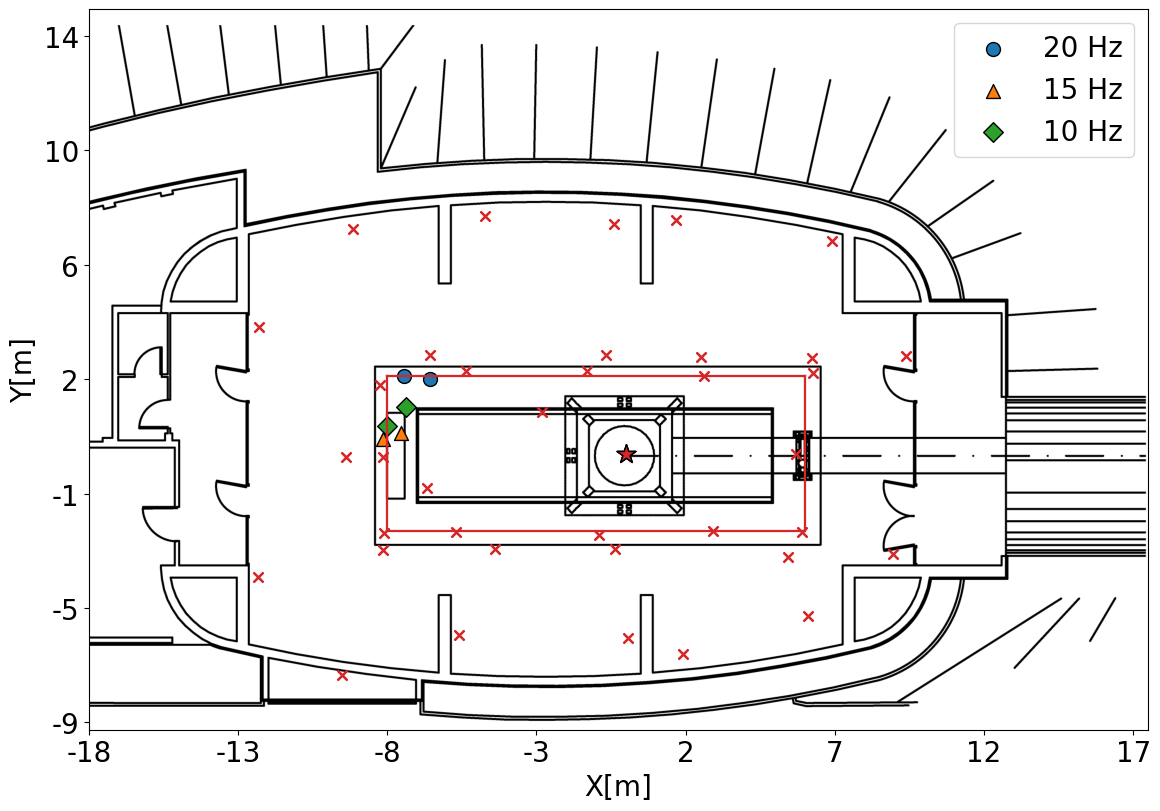}
		\includegraphics[width=0.49\textwidth]{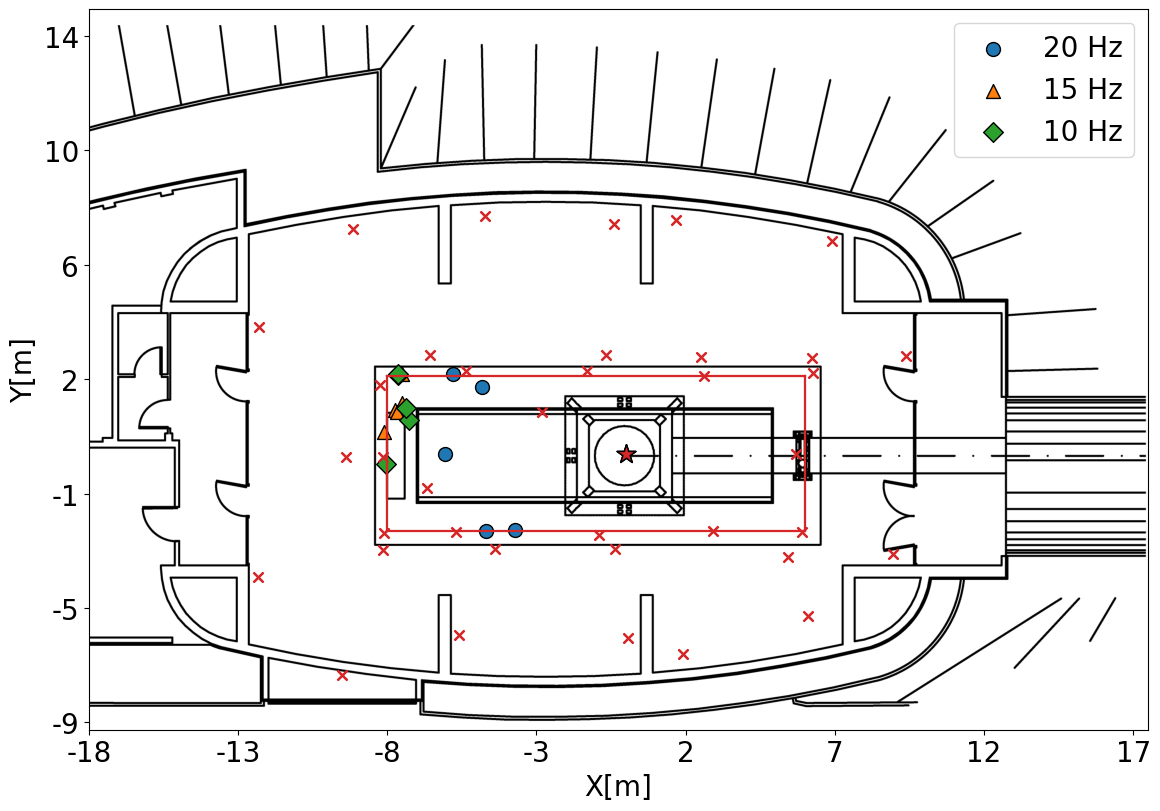}
		\includegraphics[width=0.49\textwidth]{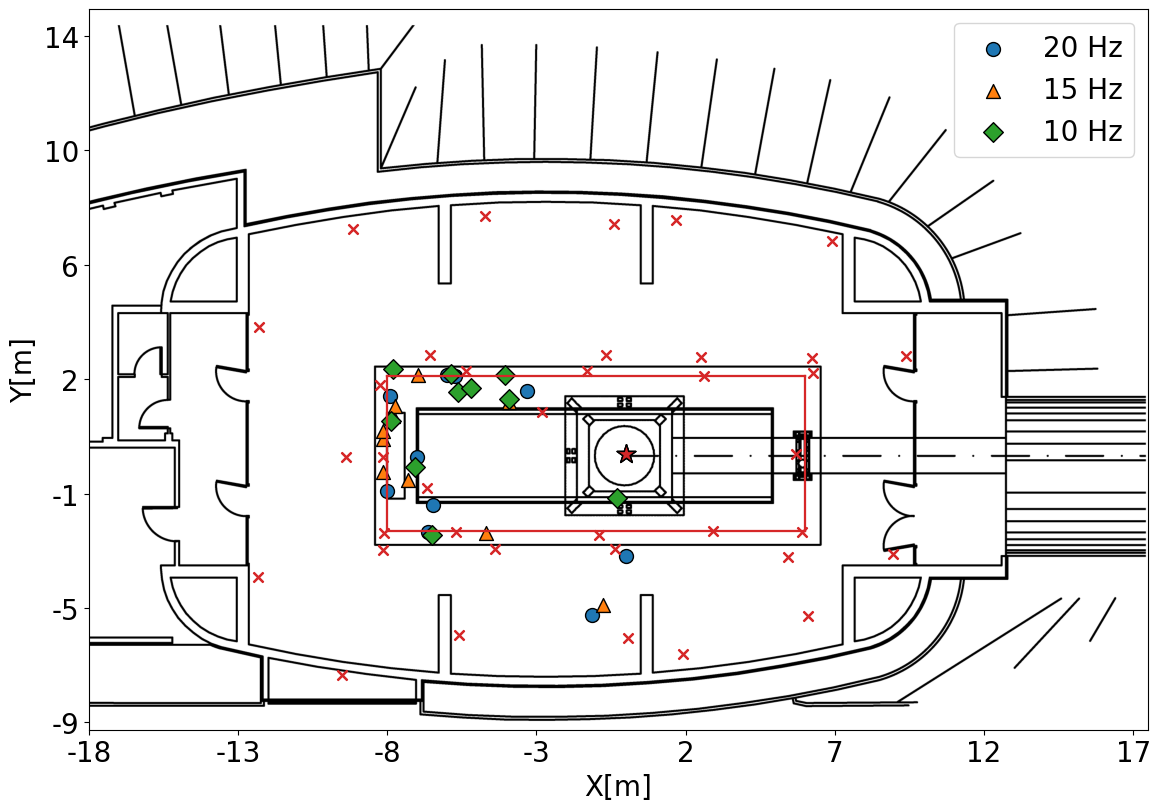}
\caption{Sample of optimal arrays with 2 (\textit{top left}), 5 (\textit{top right}), 10 (\textit{bottom}) seismometers at 10, 15 and 20 Hz}
\label{fig:optarrays}
\end{figure}

For what concerns the optimization process, we used the Particle Swarm Optimization algorithm \cite{kennedy1995particle}. Some of the results were compared with the Differential Evolution algorithm \cite{storn1997differential} to check consistency. In figure \ref{fig:optarrays}, we present results for optimizations at 10\,Hz, 15\,Hz, and 20\,Hz. Each plot contains optimal arrays with 2, 5, and 10 seismometers. It is interesting to observe that the most efficient placements of seismometers starts around the edge of the tower platform in extension of the arm, and only with 10 seismometers the array starts to occupy space closer to the test mass. This can be explained by the fact that most of the dominant seismic sources in the NN band are located in a part of the building that lies towards negative $X$ values beyond the plotted range. This means that seismic displacement is significantly stronger closer to these sources \cite{Tringali2019}. However, one might still wonder whether placing several sensors close to each other as seen in these optimal arrays is an effective strategy for NN cancellation. We will be able to explain in the following why this is the case. 

We point out a few important features of the optimized arrays. For up to 10 sensors, all sensors are located near the edge of the tower platform. It also seems very unlikely that sensors outside the building will be required for NN cancellation, which was an important open question for the design of the NN cancellation system at Virgo. The solutions also indicate that deployment of sensors on the basement floor might be advantageous. It should be noted though that only two seismometers were deployed on the basement floor (most seismometers within the red rectangle, which marks the edge of the tower platform, were deployed at surface level on the ceiling of the basement) which means that the basement contribution to the integral in equation (\ref{eq:Csn}) might be biased towards greater values. It is therefore suggested to include a larger number of basement seismometers in future measurements to make these predictions more robust.

One important observation of array optimization runs is that two optimal arrays with many more than 5 sensors are never precisely the same (albeit similar), which indicates that we might not have found the global minimum of the residual noise, but just a local minimum. This is not a problem for the design of NN cancellation systems since the Wiener filters of all arrays found by optimization have virtually the same cancellation performance. Also, we can use the results of Choromanska et al \cite{ChEA2015} to argue that the true global optimum of the array configuration does not give a significant improvement in NN cancellation compared to the solutions we present here for 10 seismometers. 

\begin{figure}[ht!]
\centering
\includegraphics[width=0.49\textwidth]{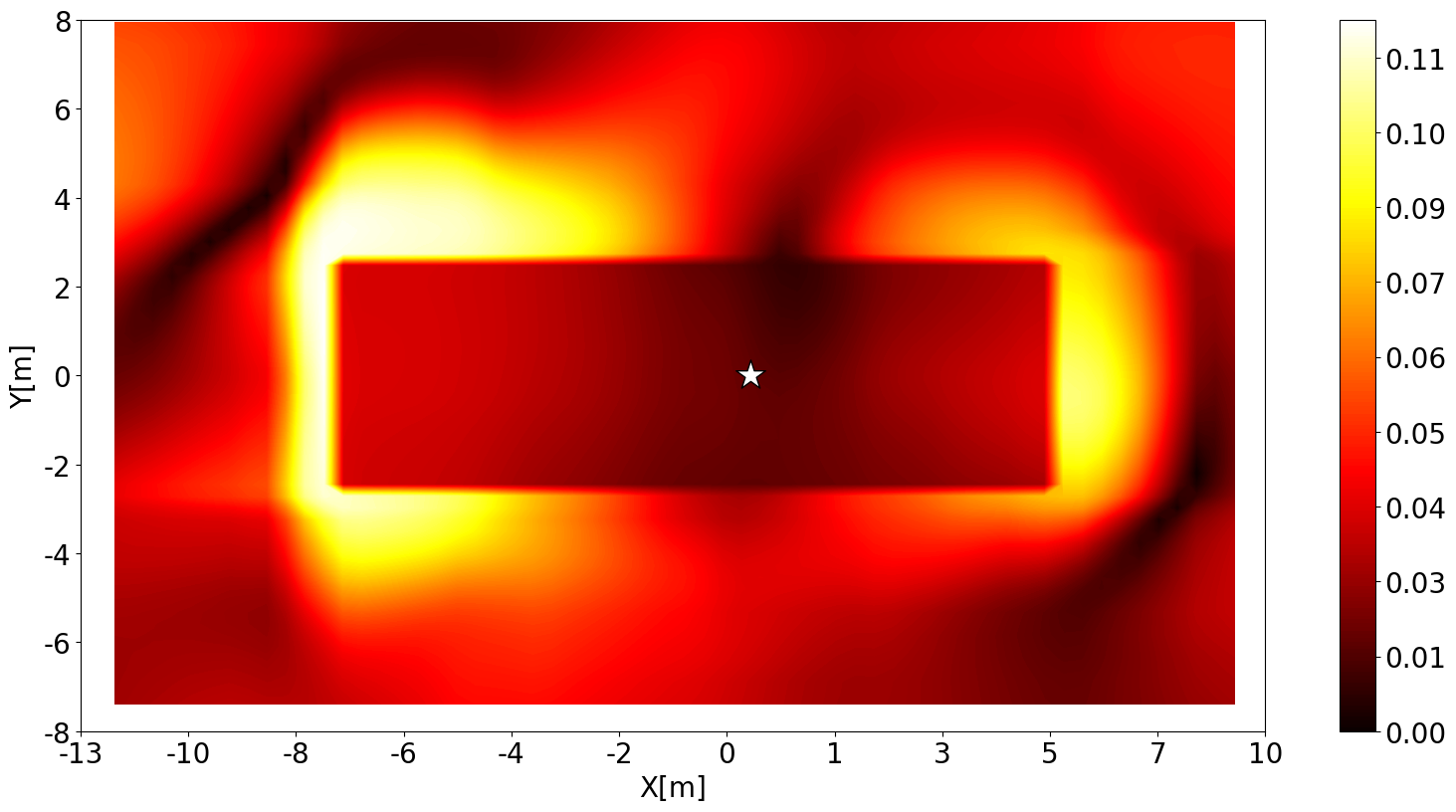}
\includegraphics[width=0.49\textwidth]{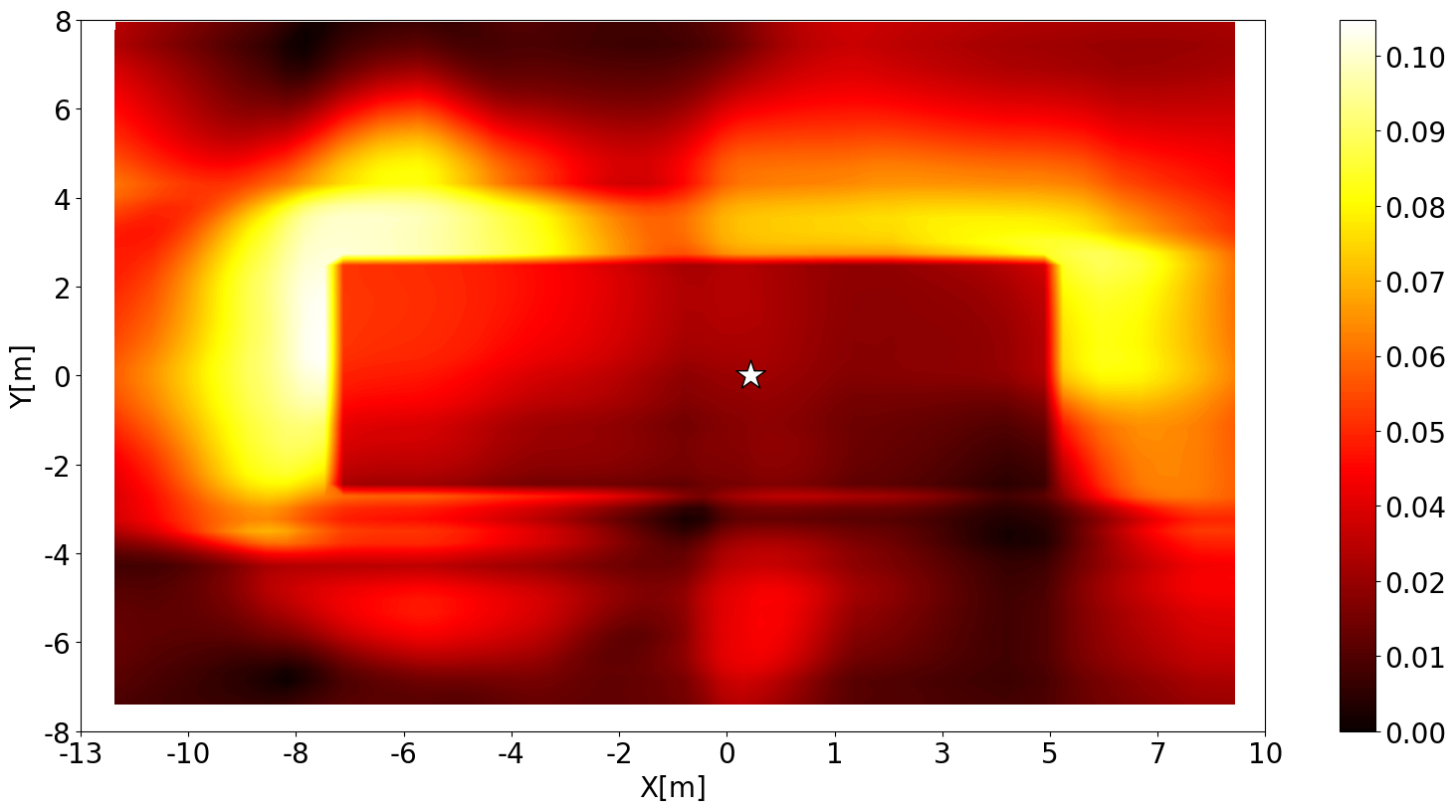}
\includegraphics[width=0.49\textwidth]{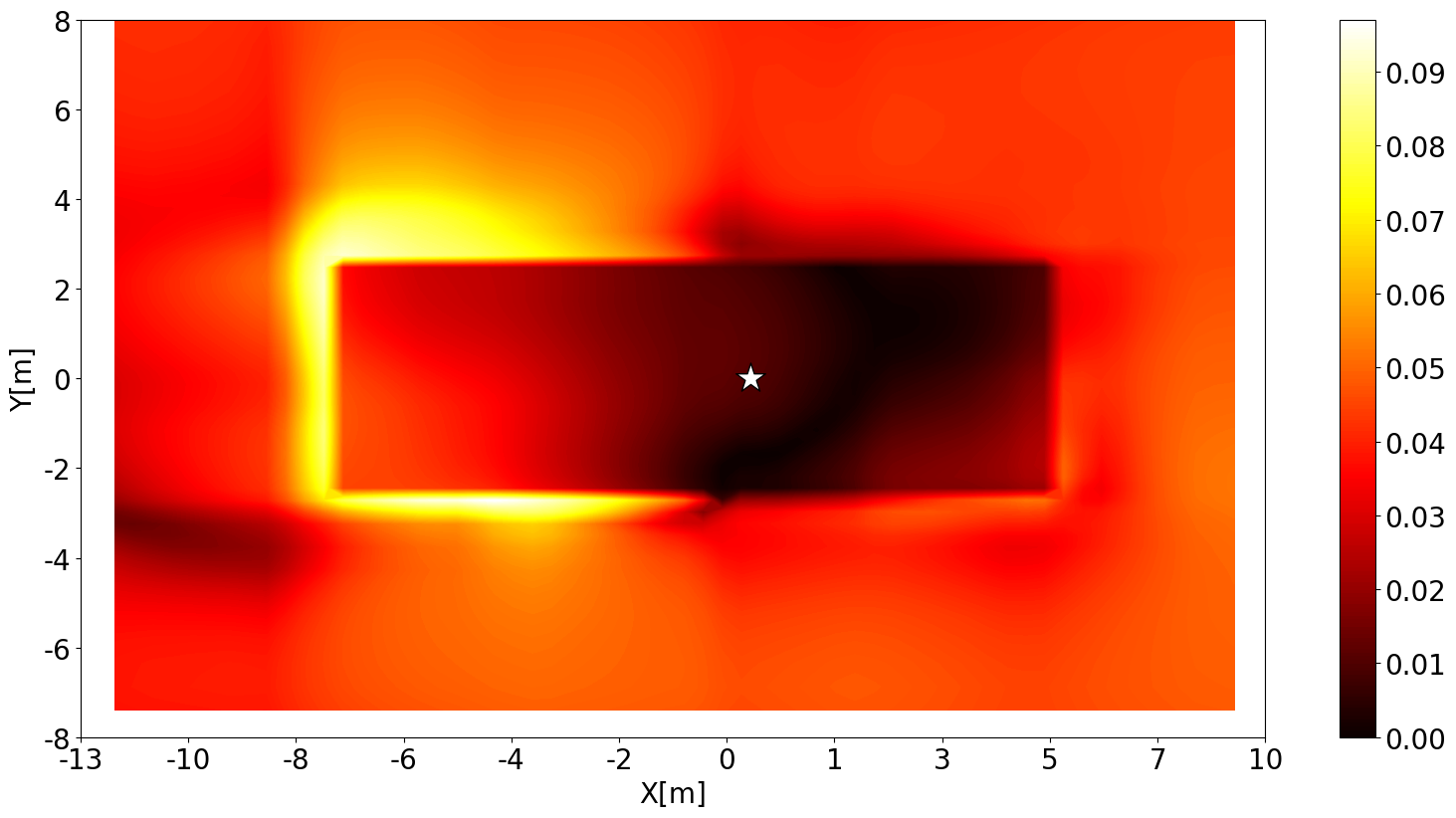}
\caption{Fraction of NN canceled ($1 - R$) by deploying a single seismometer as a function of seismometer position. \textit{top, left}: 10\,Hz, \textit{top, right}: 15\,Hz and \textit{bottom}: 20\,Hz.}
\label{fig:Csn}
\end{figure}
At first sight, the optimization results seem inconsistent with the estimated correlation $C_{\rm SN}$ of equation (\ref{eq:Csn}) shown in figure \ref{fig:Csn}. For example, the 10\,Hz plot in figure \ref{fig:Csn} shows that there is strong correlation between NN and ground motion near the edge of the tower platform towards positive $X$. Why then do optimal arrays never include sensors located there? The answer is that the seismic displacements at the two ends of the tower platform are partially correlated, which can be verified by close inspection of the two plots in figure \ref{fig:corr} for 15\,Hz. This means that placing a seismometer at one end, it is possible to cancel NN originating from seismic displacements at both ends. The negative $X$ side of the tower platform is then favored, because the part of the seismic field uncorrelated between the two ends is stronger there. This is true at 10\,Hz and 15\,Hz. In contrast, only one end of the tower platform shows significant correlation with NN at 20\,Hz. The most likely explanation for this is, as reported in \cite{Tringali2019}, that seismic waves originating from the machine rooms beyond $X=-13\,$m are reflected from the tower platform and never make it (with significant amplitude) to the other side of the tower platform. The key here is that the waves at 20\,Hz are sufficiently short to be strongly affected by the tower platform.

\begin{figure}
\centering
\includegraphics[width = 0.8\textwidth]{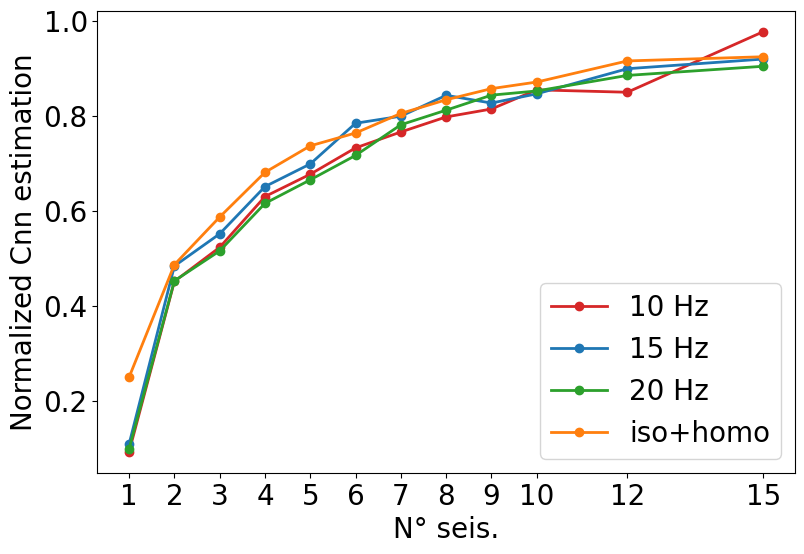}
\caption{Wiener-filter performance (1 being the maximum) as a function of number of seismometers in optimized arrays. For comparison, a theoretical performance curve is shown for the isotropic, homogeneous Rayleigh field.}
\label{fig:res}
\end{figure}

Finally, in figure \ref{fig:res}, we show the predicted Wiener-filter performance as a function of the number of seismometers in optimized arrays. More specifically, we plot a normalized version of the numerator of the last term in equation (\ref{eq:res1}). The normalization factor is the limit $N\rightarrow\infty$ of the fit 
\begin{equation}
c(N) = a\left(1-\frac{1}{bN}\right)
\label{eq:fitfunc}
\end{equation}
to the numerator in equation (\ref{eq:res1}), i.e., the value of the model parameter $a$. It is first of all surprising that all curves follow this model, which was expected to hold only for sensor-noise limited performance and to depend on properties of the seismic field \cite{BaHa2019}. Seismic correlations at 10\,Hz, 15\,Hz, and 20\,Hz are qualitatively different at the WEB, and certainly different from the correlations in an isotropic, homogeneous Rayleigh-wave field, which means that if Wiener-filter performance depends on properties of the seismic field, then the curves should look differently. This points to a yet-to-be understood universality of Wiener-filter performance in Rayleigh-wave fields, but certainly comparisons with array measurements at other sites are necessary to verify that universality holds in all cases. 

The value of $a$ in the fitting function also provides a NN estimate that takes into account the observed seismic correlations as well as the presence of the basement. These values are listed in table \ref{tab:teo-est} and compared with a theoretical model of NN in units of strain from a single test mass assuming a flat surface and isotropic, homogeneous Rayleigh field \cite{Har2019}
\begin{equation}\label{eq:iso_hom}
C_{\rm NN}^{\rm iso}(\omega) = \left(\frac{1}{L\omega^2}\right)^2\left(2\pi G\rho_0e^{-h\omega/c}\gamma\right)^2\frac{1}{2}S(\xi;\omega),
\end{equation}
where $G$ is the gravitational constant, $\rho=2500\,\rm kg/m^3$ is the density of a homogeneous medium, $\gamma=0.8$ accounts for the suppression of NN due to sub-surface (de)compression of soil by Rayleigh waves, $c=300\,$m/s is the speed of Rayleigh waves, and $S(\xi;\omega)$ is the power spectral density of vertical surface displacement. We used $h=1.5\,$m as height of test mass above ground, which would be its height if the surface at WEB were flat, i.e., without basement. Here, we have used best guesses of parameter values, since we do not have precise knowledge of average density of the ground, speed of Rayleigh waves at WEB, and the Rayleigh-NN reduction $\gamma$, which also depends on ground properties.

\begin{table}[ht!]
\renewcommand{\arraystretch}{1.3}
\begin{center}
\begin{tabular}{c|c|c}
 Frequency & $\sqrt{C_{\rm NN}^{\rm iso}}$ & $\sqrt{a}$ \\ 
 \hline
 10\,Hz & $1.33\cdot 10^{-23} \; 1/\sqrt{\rm Hz}$ & $4.04\cdot 10^{-23} \; 1/\sqrt{\rm Hz}$\\  
 15\,Hz & $7.21\cdot 10^{-24} \; 1/\sqrt{\rm Hz}$ & $1.04\cdot 10^{-23} \; 1/\sqrt{\rm Hz}$ \\
 20\,Hz & $4.34\cdot 10^{-24} \; 1/\sqrt{\rm Hz}$ & $4.44\cdot 10^{-24} \; 1/\sqrt{\rm Hz}$
\end{tabular}
\end{center}
\caption{Comparison between a theoretical model of an isotropic, homogeneous seismic field and the square root of the estimated NN PSD $a$ in equation (\ref{eq:fitfunc}).}
\label{tab:teo-est}
\end{table}

While there is a significant mismatch between our predictions and the ones obtained from a simple theoretical model, the estimated NN values $\sqrt{a}$ are in accordance with results from a finite-element simulation of an isotropic Rayleigh-wave field when including the basement (see top, right plot in figure 4 of \cite{Singha2020}). 

Now, with an estimate of $C_{\rm NN}$, we can evaluate the relative residual in equation (\ref{eq:res1}). The results are summarized in table \ref{tab:15Residual}. 
\begin{table}[ht!]
\renewcommand{\arraystretch}{1.3}
\begin{center}

\begin{tabular}{c|c|c|c}
& $\sqrt{\vec{C}^\dagger_{\rm SN}\left(\omega\right)\cdot\left(\mathbf{C}_{\rm SS}(\omega)\right)^{-1}\cdot \vec{C}_{\rm SN}(\omega)}$ & Relative residual & \makecell{ Relative residual; \\ 15\,Hz optimized}\\ 
\hline
10\,Hz & $3.95\cdot 10^{-23} \; 1/\sqrt{\rm Hz}$ & 0.02 & 0.39\\  
15\,Hz & $9.60\cdot 10^{-24} \; 1/\sqrt{\rm Hz}$ & 0.08 & 0.08\\
20\,Hz & $4.01\cdot 10^{-24} \; 1/\sqrt{\rm Hz}$ & 0.09 & 0.47
\end{tabular}
\end{center}
\caption{The second column corresponds to the PSD of the Wiener-filter output with 15 seismometers. The third column show the corresponding relative residuals $R$. The fourth column shows the relative residual achieved with an array optimized at 15 Hz.}
\label{tab:15Residual}
\end{table}
Accordingly, we predict that up to a factor 10 -- 50 reduction of NN can be achieved with an optimized array of 15 seismometers. However, one needs to keep in mind that these reductions are achieved by optimizing the array configuration at the respective frequencies. The last column shows that if we assume the same array of 15 seismometers (optimized at 15\,Hz) for the cancellation at 10\,Hz and 20\,Hz, then NN is reduced only by about a factor 2 -- 3. This might still be sufficient for Advanced Virgo+, but it is clear that we need to refine the technique if one targets a factor 10 reduction throughout the entire NN band. In fact, the current plan is to deploy around 40 seismometers around each end-test mass, and 60 at the detector vertex, which hosts the two input test masses.

\begin{figure}
\centering
\includegraphics[width = 0.8\textwidth]{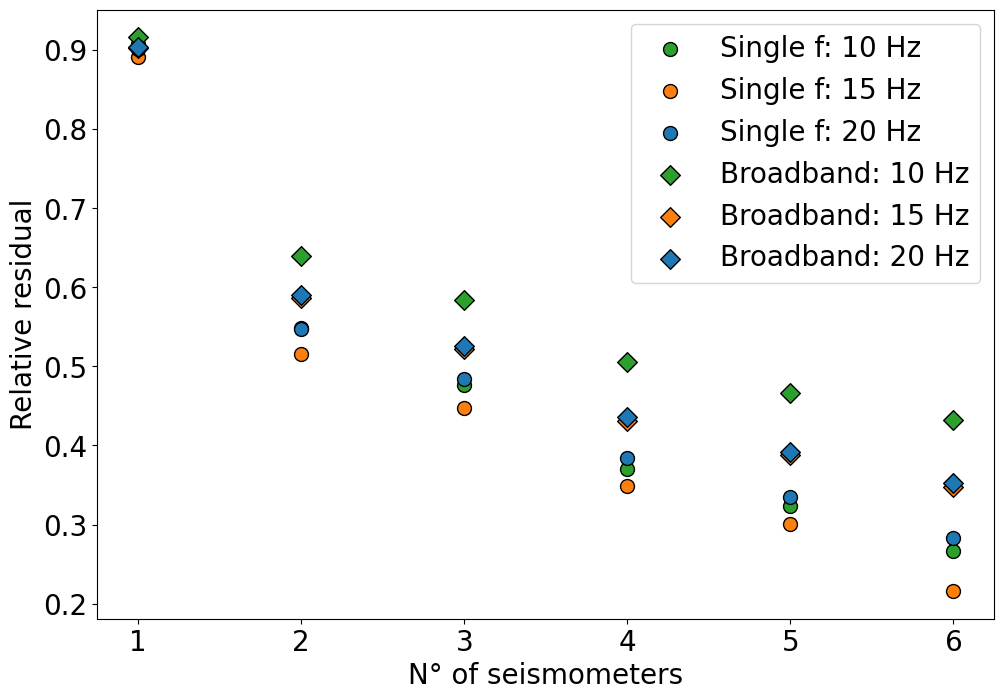}
\caption{Relative residual $R$ obtained from the broadband optimization (diamond) compared with the one obtained from the single-frequency optimization (circle).}
\label{fig:broad-res}
\end{figure}

A broadband optimization is then required. This is done by combining the single-frequency cost function of equation (\ref{eq:res1}) at different frequencies $\omega_i$. In previous work, we found that a good cost function $\mathcal L$ is given by the maximum residual \cite{BaHa2019}
\begin{equation}\label{eq:broad}
\mathcal L = \max_{\forall \omega \in \omega_i} R(\omega),
\end{equation}  
where $R(\omega)$ is calculated from equation \ref{eq:res1}, and $\omega_i$ is the set of frequencies included in the broadband optimization. We performed such an optimization using three frequencies: 10, 15 and 20\,Hz. The result was compared with the single-frequency optimization as shown in figure \ref{fig:broad-res}. Here, we first notice that, looking at a specific frequency, the broadband optimization performs worse compared to the single-frequency optimization (we achieve around 80\% of the reduction factor that we get with the single- frequency optimization). We get best cancellation performance at 15\,Hz, and in the broadband optimization, the 10\,Hz NN reduction is significantly less than at the other two frequencies.  Note that it is not surprising that one frequency has a significantly higher residual than the other two in the broadband optimization. Minimizing the broadband cost function means to find a local minimum of a single-frequency residual. The results in figure \ref{fig:broad-res} mean that there is no other local minimum of residuals at any of the three frequencies, which would leave the maximum of their residuals below the achieved 10\,Hz residual. Since the cost function in the neighborhood of the obtained optimum does not constrain the residuals at 15\,Hz and 20\,Hz, these residuals should not be expected to be at a local minimum. 

Since we do not have a theoretical model for the broadband-minimized residuals, it is difficult to extrapolate the results to higher number of sensors. Using polynomial and exponential fits, we obtain residuals of $R=0.1$ and less with at least 20 seismometers, which means a bit more than a factor 3 reduction of NN amplitude. Of course, from table \ref{tab:15Residual} we know that not more than $3\times 15$ seismometers are required to achieve $R<0.1$ at all three frequencies.
 
\section{Conclusion}
\label{sec:concl}
We have developed a surrogate Wiener filter to make best use of seismic correlation measurements at the Virgo detector for the estimation of Newtonian noise and to calculate optimal array configurations for Newtonian-noise cancellation. The approach was to use Gaussian Process Regression in combination with simple interpolation techniques. The technique is an important milestone for the design of Newtonian-noise cancellation systems of current and future GW detectors, where array configurations are to be chosen using available information of the seismic field from previous site-characterization measurements.

The method requires correlations between ground motion and GW data, which we provide by a model based on purely gravitational coupling determined by the field of seismic correlations. This model is accurate for arbitrarily complex surface displacements, but it does not consider contributions of sub-surface compression of the ground medium by body waves. As soon as Newtonian noise will be observed (or any other linear ground-to-test-mass coupling \cite{Coughlin2018,HaEA2020}), the coupling model can be substituted by correlation measurements between seismometers and GW data, which makes the optimized array configuration fully model independent.

We found that there is a universal dependence of the noise residuals on the number of seismometers used for Newtonian-noise cancellation, i.e., weakly dependent on the properties of the seismic field. Its origin should be investigated since it could be used to greatly simplify the prediction of Wiener-filter performance with optimized arrays in future detectors.

We calculated arrays with up to 15 seismometers optimized for Newtonian-noise reduction at a single frequency, which yielded a reduction by a factor 3 -- 7 in noise amplitude depending on frequency. A broadband optimization with up to 6 seismometers showed that reduction by almost a factor 2 can be achieved in amplitude. Here, extrapolation to larger numbers of sensors would not give reliable estimates since we do not have a model of the broadband residuals as a function of the number of sensors, and extrapolation depends strongly on the chosen fitting function.

\ack{The contribution of the Warsaw group was supported by the FNP grant TEAM/2016-3/19.}

\appendix

\section*{References}
\bibliography{myref}

\providecommand{\newblock}{}
\begin{thebibliography}{10}
\expandafter\ifx\csname url\endcsname\relax
  \def\url#1{{\tt #1}}\fi
\expandafter\ifx\csname urlprefix\endcsname\relax\def\urlprefix{URL }\fi
\providecommand{\eprint}[2][]{\url{#2}}

\bibitem{Har2019}
Harms J 2019 {\em Living Reviews in Relativity\/} {\bf 22} 6 ISSN 1433-8351
  \urlprefix\url{https://doi.org/10.1007/s41114-019-0022-2}

\bibitem{Sau1984}
Saulson P~R 1984 {\em Phys. Rev. D\/} {\bf 30}(4) 732--736
  \urlprefix\url{http://link.aps.org/doi/10.1103/PhysRevD.30.732}

\bibitem{HuTh1998}
Hughes S~A and Thorne K~S 1998 {\em Phys. Rev. D\/} {\bf 58}(12) 122002
  \urlprefix\url{http://link.aps.org/doi/10.1103/PhysRevD.58.122002}

\bibitem{BeEA1998}
Beccaria M, Bernardini M, Braccini S, Bradaschia C, Bozzi A, Casciano C, Cella
  G, Ciampa A, Cuoco E, Curci G, D'Ambrosio E, Dattilo V, Carolis G~D, Salvo
  R~D, Virgilio A~D, Delapierre A, Enard D, Errico A, Feng G, Ferrante I,
  Fidecaro F, Frasconi F, Gaddi A, Gennai A, Gennaro G, Giazotto A, Penna P~L,
  Losurdo G, Maggiore M, Mancini S, Palla F, Pan H~B, Paoletti F, Pasqualetti
  A, Passaquieti R, Passuello D, Poggiani R, Popolizio P, Raffaelli F,
  Rapisarda S, Vicer{\'e} A and Zhang Z 1998 {\em Classical and Quantum
  Gravity\/} {\bf 15} 3339
  \urlprefix\url{http://stacks.iop.org/0264-9381/15/i=11/a=004}

\bibitem{Cre2008}
Creighton T 2008 {\em Classical and Quantum Gravity\/} {\bf 25} 125011
  \urlprefix\url{http://stacks.iop.org/0264-9381/25/i=12/a=125011}

\bibitem{FiEA2018}
Fiorucci D, Harms J, Barsuglia M, Fiori I and Paoletti F 2018 {\em Phys. Rev.
  D\/} {\bf 97}(6) 062003
  \urlprefix\url{https://link.aps.org/doi/10.1103/PhysRevD.97.062003}

\bibitem{Singha2020}
Singha A, Hild S and Harms J 2020 {\em Classical and Quantum Gravity\/} {\bf
  37} 105007 \urlprefix\url{https://doi.org/10.1088/1361-6382/ab81cb}

\bibitem{LyEA2015}
Lynch R, Vitale S, Barsotti L, Dwyer S and Evans M 2015 {\em Phys. Rev. D\/}
  {\bf 91}(4) 044032
  \urlprefix\url{http://link.aps.org/doi/10.1103/PhysRevD.91.044032}

\bibitem{ViEv2017}
Vitale S and Evans M 2017 {\em Phys. Rev. D\/} {\bf 95}(6) 064052
  \urlprefix\url{https://link.aps.org/doi/10.1103/PhysRevD.95.064052}

\bibitem{HaEv2019}
Hall E~D and Evans M 2019 {\em Classical and Quantum Gravity\/} {\bf 36} 225002
  \urlprefix\url{https://doi.org/10.1088\%2F1361-6382\%2Fab41d6}

\bibitem{BrSe2007}
Broeck C~V~D and Sengupta A~S 2007 {\em Classical and Quantum Gravity\/} {\bf
  24} 1089--1113
  \urlprefix\url{https://doi.org/10.1088%2F0264-9381%2F24%2F5%2F005}

\bibitem{Cel2000}
Cella G 2000 {Off-Line Subtraction of Seismic Newtonian Noise} {\em Recent
  Developments in General Relativity\/} ed Casciaro B, Fortunato D,
  Francaviglia M and Masiello A (Springer Milan) pp 495--503 ISBN
  978-88-470-0068-1
  \urlprefix\url{http://dx.doi.org/10.1007/978-88-470-2113-6\_44}

\bibitem{CoEA2018a}
Coughlin M~W, Harms J, Driggers J, McManus D~J, Mukund N, Ross M~P, Slagmolen
  B~J~J and Venkateswara K 2018 {\em Phys. Rev. Lett.\/} {\bf 121}(22) 221104
  \urlprefix\url{https://link.aps.org/doi/10.1103/PhysRevLett.121.221104}

\bibitem{Tringali2019}
Tringali M~C, Bulik T, Harms J, Fiori I, Paoletti F, Singh N, Idzkowski B,
  Kutynia A, Nikliborc K, Suchi{\'{n}}ski M, Bertolini A and Koley S 2019 {\em
  Classical and Quantum Gravity\/} {\bf 37} 025005
  \urlprefix\url{https://doi.org/10.1088/1361-6382/ab5c43}

\bibitem{DHA2012}
Driggers J~C, Harms J and Adhikari R~X 2012 {\em Phys. Rev. D\/} {\bf 86}(10)
  102001 \urlprefix\url{http://link.aps.org/doi/10.1103/PhysRevD.86.102001}

\bibitem{CoEA2016a}
Coughlin M, Mukund N, Harms J, Driggers J, Adhikari R and Mitra S 2016 {\em
  Classical and Quantum Gravity\/} {\bf 33} 244001
  \urlprefix\url{http://stacks.iop.org/0264-9381/33/i=24/a=244001}

\bibitem{BHC2008}
Benesty J, Huang Y and Chen J 2008 {Wiener and Adaptive Filters} {\em Springer
  Handbook of Speech Processing\/} ed Benesty J, Sondhi M and Huang Y (Springer
  Berlin Heidelberg) pp 103--120 ISBN 978-3-540-49125-5
  \urlprefix\url{http://dx.doi.org/10.1007/978-3-540-49127-9_6}

\bibitem{HaVe2016}
Harms J and Venkateswara K 2016 {\em Classical and Quantum Gravity\/} {\bf 33}
  234001 \urlprefix\url{http://stacks.iop.org/0264-9381/33/i=23/a=234001}

\bibitem{BaHa2019}
Badaracco F and Harms J 2019 {\em Classical and Quantum Gravity\/} {\bf 36}
  145006 \urlprefix\url{https://doi.org/10.1088%2F1361-6382%2Fab28c1}

\bibitem{jones2001taxonomy}
Jones D~R 2001 {\em Journal of global optimization\/} {\bf 21} 345--383

\bibitem{forrester2008engineering}
Forrester A, Sobester A and Keane A 2008 {\em Engineering design via surrogate
  modelling: a practical guide\/} (John Wiley \& Sons)

\bibitem{press2007numerical}
Press W~H, Teukolsky S~A, Vetterling W~T and Flannery B~P 2007 {\em Numerical
  recipes 3rd edition: The art of scientific computing\/} (Cambridge university
  press)

\bibitem{kennedy1995particle}
Kennedy J and Eberhart R 1995 Particle swarm optimization {\em Proceedings of
  ICNN'95-International Conference on Neural Networks\/} vol~4 (IEEE) pp
  1942--1948

\bibitem{storn1997differential}
Storn R and Price K 1997 {\em Journal of global optimization\/} {\bf 11}
  341--359

\bibitem{ChEA2015}
Choromanska A, Henaff M, Mathieu M, {Ben Arous} G and LeCun Y 2015 {\em Journal
  of Machine Learning Research\/} {\bf 38} 192--204 ISSN 1532-4435

\bibitem{Coughlin2018}
Coughlin M, Harms J, Driggers J, McManus D, Mukund N, Ross M, Slagmolen B and
  Venkateswara K 2018 {\em Physical Review Letters\/} {\bf 121}
  \urlprefix\url{https://doi.org/10.1103/physrevlett.121.221104}

\bibitem{HaEA2020}
Harms J, Bonilla E~L, Coughlin M~W, Driggers J, Dwyer S~E, McManus D~J, Ross
  M~P, Slagmolen B~J~J and Venkateswara K 2020 {\em Phys. Rev. D\/} {\bf
  101}(10) 102002
  \urlprefix\url{https://link.aps.org/doi/10.1103/PhysRevD.101.102002}

\end{thebibliography}
\bibliographystyle{iopart-num}

\end{document}